\documentclass[epj]{webofc}
\usepackage[varg]{txfonts}   % Web of Conferences font
\woctitle{MESON2016 - the 14$^\textrm{th}$ International Workshop on Meson Production, Properties and Interaction}
\begin{document}
\selectlanguage{english}
\title{Production and interaction of the $\eta$ meson with nucleons and nuclei}

% insert email only for speaker/presenter
\author{Wojciech Krzemie\'n\inst{1}\fnsep\thanks{\email{wojciech.krzemien@ncbj.gov.pl}} \and
  Oleksandr Khreptak~\inst{2} \and
  Pawe\l~Moskal~\inst{2} \and
  Iryna Ozyrianska-Sch\"atti~\inst{2} \and
  Oleksandr Rundel~\inst{2} \and
  Magdalena Skurzok~\inst{2} \and
  Marcin Zieli\'nski\inst{2}
% comment out the next line if not needed
       \\for the WASA-at-COSY Collaboration
}

\institute{National Centre for Nuclear Research, 05-400 Otwock, Świerk, Poland
\and
          Faculty of Physics, Astronomy and Computer Science, Jagiellonian University, 30-348 Cracow, Poland}

\abstract{%Do not break line here!
We report on the status of the search for $\eta$-mesic nuclei and the studies of the interaction of the $\eta$ meson with nucleons. Recently we have completed the analysis of the new WASA-at-COSY data on the production of the $\eta$ meson with polarized proton beam. New results on the analyzing power for the $\vec{p}p\rightarrow pp\eta$ reaction with more than an order of magnitude improved precision shed a new light on the production mechanism of the $\eta$ meson in nucleon-nucleon collisions. Also, the latest results of the search for $\eta$-mesic nuclei are discussed.
}
\maketitle
\section{Introduction}
\label{intro}
%The $\eta$ particle~\cite{PDG}, has several
%properties which makes it interesting to study. 
The $\eta$ particle together with isoscalar $\eta'$ and isovector $\pi^0$ lay in the origin ($S=0$, $I_3=0$) of  the nonet of pseudoscalar mesons 
representation. However, its behaviour is very different with respect to its interaction with nucleons~\cite{Mosk1}.
In the low energy region, 
the $\eta$ meson interaction with nucleons is dominated by the $S_{11}$ resonance, which with
its mass of 1535 MeV lays very close to the $\eta$-N threshold. 
This makes the s-wave $\eta-N$ interaction very strong and - as shown the analysis of Bhalerao and Liu - attractive~\cite{Bhalerao}. This can be contrasted with the pion case which, dominated by the p-wave interaction from the $\Delta(1232)$ resonance, is much weaker~\cite{Krusche_Wilkin}.
Also, the measurement of the $\eta'$N scattering length shows that its interaction is rather weak~\cite{Eryk}.
 The large value of the $\eta$-N scattering length led to the hypothesis, proposed by the Haider and Liu, who postulated that the total interaction in a nucleus- system is strong enough to form a bound-state  - the so called mesic nuclei~\cite{HaiderLiu1}.
The second question raised and not unequivocally answered by the earlier measurements was about the $\eta$ production mechanism 
in the nucleon-nucleon collisions.

Due to the short lifetime of the  meson ($t \approx 10^{-18}$ s) it is not feasible to
create the $\eta$ beam. Therefore, its interaction with nucleon or nuclei must
be studied via the observation of final states of nuclear reactions including
the $\eta$-nucleon (or $\eta$-nuclei) pair. 
The Final State Interaction between produced particles can strongly 
influence the production cross-sections and, in this way, can be used for studies of the interaction
itself~\cite{Mosk2}.

In this contribution we discuss the $\eta$ production mechanism in the interaction with nucleons 
and the search for the $\eta$-mesic nuclei in the context of the recent experimental results from
WASA-at-COSY collaboration.

\section{$\eta$ production mechanism in the interaction with nucleons } 
The measurements of the large total cross-section of $NN \rightarrow NN \eta$ reaction
near the $\eta$ production threshold~\cite{Chiavasa,Calen1,Calen2,Hibou,Smyrski,Bergdolt,Abdel,MoskalP1, MoskalP2,Petren,Calen3, Mosk3} motivated the two-step $\eta$ production
model proposed in~\cite{Faldt}. In this scenario one of the proton is firstly excited through
the exchange of a single meson and forms the $S_{11}$ resonance, which in the second step deexcites via the emission of $\eta$ and nucleon.
In principle, the excitation to $S_{11}$ state can occur by exchanging $\pi$, $\eta$,
$\omega$ or/and $\rho$ mesons. The measurements of total cross-section isospin dependence 
by WASA/PROMICE and COSY-11 showed that the $\eta$ production in the total isosinglet state is much 
higher that in the isotriplet state~\cite{promice,Mosk3}. This result strongly suggest the isovector
meson exchange, reducing the candidates to $\pi$ and $\rho$ particles~\cite{Nakayama1, Nakayama2, Faldt}. 
To further distinguish  between the $\pi$ and $\rho$ meson exchange models, 
the determination of the analysing power in the polarization
measurement was required. The first measurement by COSY-11 gave a indication in favour
of the pseudoscalar exchange model,  
although due to the limited statistics the decisive conclusions could not be done~\cite{Winter1, Winter2, Czyzyk}.
The WASA-at-COSY performed a high-statistics measurement of the
$\vec{p}p \rightarrow pp\eta$ with the polarized beam. The spin of the polarization was
flipped from cycle to cycle. The data was gathered for two separated beam momenta
2026 MeV/c and 2188 MeV/c, which correspond to excess energy over the $\eta$ production threshold
of 15 MeV and 72 MeV, respectively. More details of the analysis can be found in~\cite{Iryna}.

\section{Search for the $\eta$-mesic nuclei} 
The recent reviews on the search for mesic nuclei can be found in~\cite{Gal, Machner_2015, Wilkin_2016, Krusche_Wilkin, Kelkar, SBass, HaiderLiu,Moskal_FewBody}. 
The WASA-at-COSY collaboration~\cite{WASA_dsc} performed three dedicated experiments with aim to search
for the $\eta$-mesic nuclei in ${^4\mbox{He}}$ and ${^3\mbox{He}}$ systems in the deuteron-deuteron
and proton-deuteron collisions, respectively~\cite{Adlarson_2013, WKrzemien_2014, Acta_2016,Krzemien_PhD, Skurzok_PhD}.
The choice of the light nuclei was motivated by both theoretical considerations
(see e.g.~\cite{WycechGreen, Wilkin1}) as well as the earlier measurements by SATURNE, ANKE and COSY-11~\cite{Willis97, moskalsymposium, jurek-he3, timo, meson08}, that provided strong experimental hints
for the existence of the bound state in the $^3\mbox{He}-\eta$ and $^4\mbox{He}-\eta$ systems.
The main experimental idea for the ${^4\mbox{He}}$ is based on the measurement of the excitation function
of the $dd \rightarrow {^3\mbox{He}} N \pi$ reaction for energies in the vicinity of the $\eta$ production
threshold and on the selection of events with low ${^3\mbox{He}}$ center-of-mass (CM) momenta.
In the case of existence of the ${^4\mbox{He}}-\eta$ bound state we expect to observe
a resonance-like structure in the excitation function below the threshold for the production of the ${^4\mbox{He}}-\eta$ system.
The ${^3\mbox{He}}$ state is investigated in proton on deuteron collisions. The details can be found in~\cite{Rundel}.

\section{Summary}
The latest preliminary results from the 2010 experiment in $^4\mbox{He}-\eta$ system, do not confirm the existence of the
$\eta$-mesic nuclei. 
The preliminary value of the upper limit obtained from the simultaneous fit, taking into account the isospin dependence  of the 
$dd\rightarrow(^{4}\hspace{-0.03cm}\mbox{He}$-$\eta)_{bound}\rightarrow$ $^{3}\hspace{-0.03cm}\mbox{He} n \pi{}^{0}$ and  $dd\rightarrow(^{4}\hspace{-0.03cm}\mbox{He}$-$\eta)_{bound}\rightarrow$ $^{3}\hspace{-0.03cm}\mbox{He} p \pi{}^{-}$ excitation functions
, of order of few nb can be compared to the theoretical estimate of 4 nb~\cite{WycechKrzemien}. 
In case of the $^3\mbox{He}-\eta$ system, the analysis is ongoing. The current experimental upper limit for the
production of $pd \rightarrow$ ($^{3}\hspace{-0.03cm}\mbox{He}$-$\eta)_{bound} \rightarrow$ $p p p \pi{}^{-}$  comes from the COSY-11 measurement and is equal to about 270 nb~\cite{MosSmy}. Due to the high statistics
the expected sensitivity in current WASA analysis is of order of 10 nb, which, taking into account the
theoretical estimate of 80 nb~\cite{Wilkin_Acta2014}, should be sufficient to confirm or rule out the hypothesis of existence of the $^3\mbox{He}-\eta$ mesic nuclei.

The WASA-at-COSY determined the analysing power in the $\vec{p}p \rightarrow pp\eta$ reaction with two order of magnitude
higher precision than the previous COSY-11 measurement.
The preliminary results of the angular dependency of the analysing power is in disagreement with the prediction by both the pseudo-scalar and the vector exchange models. For higher energy (Q =72 MeV), the Ps-Pp interference is clearly observed.

\begin{acknowledgement}
We acknowledge support by the Polish National Science Center through grant
No. 2011/03/B/ST2/01847,2011/01/B/ST2/00431, 2013/11/N/ST2/04152,
by the FFE grants of the Research Center Juelich, by the EU Integrated
Infrastructure Initiative HadronPhysics Project under contract number RII3-CT-2004-506078
and by the European Commission under the 7th Framework Programme through the Research
Infrastructures action of the Capacities Programme, Call: FP7- INFRASTRUCTURES-2008-1,
Grant Agreement N. 227431.
\end{acknowledgement}
%
% BibTeX or Biber users please use (the style is already called in the class, ensure that the "woc.bst" style is in your local directory)
% \bibliography{name or your bibliography database}
%
% Non-BibTeX users please use
%

\end{document}